\documentclass[aps,prd,preprintnumbers,superscriptaddress,nofootinbib,notitlepage]{revtex4-1}
\usepackage[dvipdfmx]{graphicx}
\usepackage{bm,latexsym,amsmath,amssymb,amsfonts,mathrsfs}
\usepackage{color}
\input{colordvi.tex}
\allowdisplaybreaks[1]
\newcommand*{\D}{{\rm d}}
\newcommand*{\mpl}{M_{\rm Pl}}
\begin{document}
%-----------------------------------------------------------------%
\title{On the gauge dependence of gravitational waves generated at %
second order from scalar perturbations}
%-----------------------------------------------------------------%

\author{Keitaro~Tomikawa}
\email[Email: ]{k.tomikawa"at"rikkyo.ac.jp}
\affiliation{Department of Physics, Rikkyo University, Toshima, Tokyo 171-8501, Japan
}

\author{Tsutomu~Kobayashi}
\email[Email: ]{tsutomu"at"rikkyo.ac.jp}
\affiliation{Department of Physics, Rikkyo University, Toshima, Tokyo 171-8501, Japan
}

%-----------------------------------------------------------%
\begin{abstract}
We revisit and clarify the gauge dependence of gravitational waves generated
at second order from scalar perturbations.
In a universe dominated by a perfect fluid with
a constant equation-of-state parameter $w$,
we compute the energy density of such induced gravitational waves
in the Newtonian, comoving, and uniform curvature gauges.
Huge differences are found between the Newtonian and comoving gauge results
for any $w \,(\ge 0)$.
This is always caused by the perturbation of the shift vector.
Interestingly, the Newtonian and uniform curvature gauge
calculations give the same energy density for $w>0$.
In the case of $w=0$,
the uniform curvature gauge result
differs only by a factor from that of the comoving gauge,
but deviates significantly from that of the Newtonian gauge.
Our calculation is done analytically for $w=0$ and $w=1/3$,
and our result is consistent with the previous numerical one.
\end{abstract}
%-----------------------------------------------------------------%
%\pacs{%}
\preprint{RUP-19-27}
\maketitle
%-----------------------------------------------------------------%

\section{Introduction}

After the detection of gravitational waves (GWs) by the LIGO
and Virgo collaborations~\cite{Abbott:2016blz,Abbott:2016nmj,Abbott:2017vtc,%
Abbott:2017oio,TheLIGOScientific:2017qsa,Abbott:2017gyy},
it is becoming more and more important to study
GWs generated from various sources.
Among a variety of sources, scalar (density) perturbations
at quadratic order~\cite{Mollerach:2003nq,Ananda:2006af,Baumann:2007zm} are of
particular interest.
Though the detectability of the GWs induced at second order
depends on the cosmological scenarios under consideration, at least
we know that their sources, i.e., scalar perturbations, do exist in the Universe.
Since the energy-density spectrum of induced GWs is determined from
the cosmic expansion history and
the primordial spectrum of scalar perturbations,
it can be a powerful probe for different scenarios
having particular features in these respects~\cite{Saito:2008jc,%
Assadullahi:2009nf,Saito:2009jt,Bugaev:2009zh,%
Jedamzik:2010hq,Alabidi:2012ex,Alabidi:2013wtp,Inomata:2018epa,%
Inomata:2019zqy,Inomata:2019ivs,Cai:2018dig,Cai:2019cdl}.

Almost all previous studies on scalar-induced GWs
have employed the Newtonian gauge for the scalar perturbations at linear order.
At second order in cosmological perturbation theory, however,
tensor perturbations are dependent on the gauge choice of
the scalar perturbations, as emphasized earlier in Ref.~\cite{Arroja:2009sh}.
This is in contrast to the tensor perturbations at linear order.
One would thus notice that there is no a priori reason for using only the Newtonian gauge.
The source term for induced GWs during the matter-dominated era
was obtained in the comoving gauge in Ref.~\cite{Boubekeur:2008kn},
but explicit solutions for the induced GWs have not been
discussed in depth there.
More recently, the explicit calculation of induced GWs
in different gauges has been presented in Ref.~\cite{Hwang:2017oxa},
where, interestingly enough, a significant gauge dependence has been reported.
The main result of Ref.~\cite{Hwang:2017oxa}
was obtained numerically, using the standard cosmological model
with the best-fitting cosmological parameters.
Here the following questions would arise:
\begin{itemize}
  \item Can we understand this gauge dependence more analytically?
  \item How does this gauge dependence depend on the background cosmological evolution
  (more specifically, the equation-of-state parameter)?
  \item How does the gauge dependence depend on the input form of the
  primordial power spectrum of the scalar perturbations?
\end{itemize}
To clarify these points, we consider a universe filled with
a single perfect fluid with a constant equation-of-state parameter,
and evaluate the energy density of induced GWs
in three representative gauges:
the Newtonian gauge, the comoving gauge, and the
uniform curvature gauge.
Our calculation is done analytically for
matter-dominated and radiation-dominated universes
following Ref.~\cite{Kohri:2018awv} and numerically in the other cases.

This paper is organized as follows.
In the next section,
we derive the basic formula for the energy density of induced GWs.
In Sec.~III, we introduce the action approach to GWs induced from
scalar perturbations to derive the gauge-ready form of the source term.
Then, in Sec.~IV we discuss the gauge dependence of induced GWs
by evaluating their evolution analytically and numerically for
different values of the equation-of-state parameter.
Our conclusion is drawn in Sec.~V.

\section{Induced Gravitational Waves}\label{App:Omega GW}

We start with deriving the expression for the energy density
of induced GWs.
The evolution equation for the induced GWs $h_{ij}(\eta,\Vec{x})$
is of the form
\begin{align}
h_{ij}''+2{\cal H}h_{ij}'-\partial^2h_{ij}=\Lambda_{ij,kl}S_{kl},
\label{EoM of h genuniversel}
\end{align}
where a dash denotes differentiation with respect to the
conformal time $\eta$ and
$\Lambda_{ij,kl}$ is the projection tensor which extracts
the transverse-traceless part of the source term $S_{kl}(\eta,\Vec{x})$.
A more explicit expression for $\Lambda_{ij,kl}S_{kl}$
will be given shortly below.
We consider a universe filled with a perfect fluid
whose equation of state parameter $w$ is a constant.
The scale factor is then given by $a\propto \eta^{2/(1+3w)}$
and hence ${\cal H}:=a'/a=[2/(1+3w)]/\eta$.

The Fourier components of $h_{ij}$ are defined by
\begin{align}
h_{ij}(\eta,\Vec{x})=
\sum_{A}\int \frac{\D^3k}{(2\pi)^{3/2}}
h_A(\eta,\Vec{k})e^{(A)}_{ij}(\Vec{k})e^{i\Vec{k}\cdot\Vec{x}}\qquad (A=+,\times).
\end{align}
The two polarization tensors $e^{(A)}_{ij}(\Vec{k})$ are defined as
\begin{align}
    e^{(+)}_{ij}(\Vec{k})&=\frac{1}{\sqrt{2}}\left [ e_{i}(\Vec{k})e_{j}(\Vec{k})-\overline{e}_{i}(\Vec{k})\overline{e}_{j}(\Vec{k}) \right], \\
    e^{(\times)}_{ij}(\Vec{k})&=\frac{1}{\sqrt{2}}\left[ e_{i}(\Vec{k})\overline{e}_{j}(\Vec{k})+\overline{e}_{i}(\Vec{k})e_{j}(\Vec{k}) \right],
\end{align}
where $e_{i}(\Vec{k})$ and $\overline e_{j}(\Vec{k})$
are unit vectors orthogonal to each other and to $\Vec{k}$.
It follows from the definition of the polarization tensors that
$k^ie_{ij}^{(A)}=0$ and $e_{ij}^{(A)}(\Vec{k})e_{ij}^{(A')}(\Vec{k})=\delta_{AA'}$.

Using the polarization tensors one can write
the right-hand side of Eq.~\eqref{EoM of h genuniversel} as
\begin{align}
\Lambda_{ij,kl}S_{kl}(\eta,\Vec{x})
=\sum_{A}\int \frac{\D^3k}{(2\pi)^{3/2}}
e_{ij}^{(A)}(\Vec{k})e_{lm}^{(A)}(\Vec{k})S_{lm}(\eta,\Vec{k}),
\end{align}
where
\begin{align}
S_{ij}(\eta,\Vec{k}):=
\int \frac{\D^3x}{(2\pi)^{3/2}}
S_{ij}(\eta,\Vec{x})e^{-i\Vec{k}\cdot\Vec{x}}
\end{align}
is the Fourier transform of $S_{ij}(\eta,\Vec{x})$.
We write $S_{A}(\eta,\Vec{k}):=e_{ij}^{(A)}(\Vec{k})S_{ij}(\eta,\Vec{k})$.

The formal solution to Eq.~\eqref{EoM of h genuniversel} in the Fourier domain
is given by
\begin{align}
h_{A}(\eta,\Vec{k})=
\frac{1}{a(\eta)}\int^\eta_0G_k(\eta,\bar{\eta})a(\bar\eta)
S_{A}(\bar\eta,\Vec{k})\D\bar\eta,
\end{align}
where the Green's function is
expressed in terms of the Bessel functions as
\begin{align}
G_k(\eta,\bar\eta)=\frac{\pi}{2}\eta^{1/2}\bar\eta^{1/2}\left[
Y_\nu(k\eta)J_\nu(k\bar \eta)-J_\nu(k\eta)Y_\nu(k\bar \eta)
\right],
\end{align}
with
\begin{align}
\nu:=\frac{3(1-w)}{2(1+3w)}.
\end{align}

The energy density of GWs is given by
\begin{align}
\rho_{\rm GW}(\eta)=\frac{\mpl^2}{2}\sum_A\frac{1}{a^2}\langle
h_A'(\eta,\Vec{x})h_A'(\eta,\Vec{x})
\rangle,
\end{align}
where $\langle\cdots\rangle$ denotes a spatial average and
\begin{align}
h_A(\eta,\Vec{x}):=\int\frac{\D^3k}{(2\pi)^{3/2}}h_A(\eta,\Vec{k})e^{i\Vec{k}\cdot\Vec{x}}.
\end{align}
Since
\begin{align}
  \langle
  h_A'(\eta,\Vec{x})h_A'(\eta,\Vec{x})
  \rangle
  =\int \frac{\D^3k}{(2\pi)^{3/2}}\frac{\D^3k'}{(2\pi)^{3/2}}
  h_A'(\eta,\Vec{k}){h^*_A}'(\eta,\Vec{k}')\langle e^{i(\Vec{k}-\Vec{k}')\cdot\Vec{x}}
  \rangle,
\end{align}
and
\begin{align}
  \langle e^{i(\Vec{k}-\Vec{k}')\cdot\Vec{x}}
  \rangle=\frac{1}{V}\int e^{i(\Vec{k}-\Vec{k}')\cdot\Vec{x}}\D^3 x
  =\frac{(2\pi)^3}{V}\delta^{(3)}(\Vec{k}-\Vec{k}'),
\end{align}
where $V$ is a volume whose size is much larger than the wavelengths of
interest,
we have
\begin{align}
\rho_{\rm GW}(\eta)
=\frac{\mpl^2}{2}\frac{1}{Va^2}\sum_A \int \D^3k
|h_A'(\eta,\Vec{k})|^2.
\end{align}
In the subhorizon regime, $k \eta \gg 1$,
the time derivative of $h_{A}$ is approximately given by
\begin{align}
h_A'(\eta,\Vec{k})\simeq \frac{1}{a(\eta)}\int^\eta_0 \partial_\eta
G_k(\eta,\bar\eta)a(\bar\eta)S_{A}(\bar\eta,\Vec{k})\D\bar\eta.
\end{align}
We find the following approximate expression for the
time derivative of the Green's function,
\begin{align}
  \partial_\eta
  G_k(\eta,\bar\eta)\simeq {\cal G}_k(\eta,\bar\eta):=  \frac{\pi}{2}(k\eta)^{1/2}
  (k\bar\eta)^{1/2}\left[
  Y_{\nu-1}(k\eta)J_\nu(k\bar \eta)-J_{\nu-1}(k\eta)Y_\nu(k\bar \eta)
  \right].
\end{align}
More explicitly, for $\nu=1/2$ ($w=1/3$) we have
\begin{align}
{\cal G}_k(\eta,\bar\eta)=\cos[k(\eta-\bar\eta)],
\end{align}
and
for $\nu=3/2$ ($w=0$) we have
\begin{align}
{\cal G}_k(\eta,\bar\eta)=\cos[k(\eta-\bar\eta)]+\frac{\sin[k(\eta-\bar\eta)]}{k\bar\eta}.
\end{align}
It then follows that
\begin{align}
\rho_{\rm GW}(\eta)=\frac{\mpl^2}{2}\frac{1}{Va^4(\eta)}\sum_A
\int\D^3k \int^\eta_0\D\eta'\int^\eta_0\D\eta''a(\eta')a(\eta'')
{\cal G}_k(\eta,\eta'){\cal G}_k(\eta,\eta'')S_A(\eta',\Vec{k})S_A^*(\eta'',\Vec{k}).
\label{end_gw_1}
\end{align}

We are interested in the case where
$S_A$ is of the form
\begin{align}
S_A(\eta, \Vec{k})=e^{(A)}_{ij}(\Vec{k})\int \frac{\D^3q}{(2\pi)^{3/2}}
q^iq^jA(\Vec{q})A(\Vec{k}-\Vec{q})F(\Vec{k},\Vec{q},\eta),
\label{def:F}
\end{align}
where $A(\Vec{q})$ is a Gaussian random field and
$F(\Vec{k},\Vec{q},\eta)$ is some function.
In the actual calculation, $A(\Vec{q})$ will be the primordial amplitude of
scalar perturbations and $F$ contains the information
of their time evolution.
The power spectrum ${\cal P}(k)$ is defined by
\begin{align}
\langle A(\Vec{k})A^*(\Vec{q})\rangle =\frac{2\pi^2}{k^3}
\delta^{(3)}(\Vec{k}-\Vec{q}){\cal P}(k).
\end{align}
(Now $\langle\cdots\rangle$ denotes an average over the whole distribution.)
Using Wick's theorem,
the two-point correlator of the source term $S_A$
can be written as
\begin{align}
\langle S_A(\eta',\Vec{k})S_A^*(\eta'',\Vec{k}')\rangle
&=\int\frac{\D^3q}{(2\pi)^3}
(e_{ij}^{(A)}(\Vec{k})q^iq^j)^2
\frac{8\pi^4}{q^3|\Vec{k}-\Vec{q}|^3}
\delta^{(3)}(\Vec{k}-\Vec{k}'){\cal P}(q){\cal P}(|\Vec{k}-\Vec{q}|)
F(\Vec{k},\Vec{q},\eta')
F(\Vec{k},\Vec{q},\eta'').
\end{align}
Using this and
$[\delta^{(3)}(\Vec{k}-\Vec{k}')]^2=[V/(2\pi)^3]\delta^{(3)}(\Vec{k}-\Vec{k}')$,
one can write the energy density of GWs~\eqref{end_gw_1} as
\begin{align}
\rho_{\rm GW}(\eta)&=\frac{2\mpl^2}{a^4(\eta)}
\int\frac{\D^3k}{(2\pi)^3}
\int\frac{\D^3q}{(2\pi)^3}\int^\eta_0\D\eta'\int^\eta_0\D\eta''
a(\eta')a(\eta''){\cal G}_k(\eta,\eta'){\cal G}_k(\eta, \eta'')
\notag \\ &\quad \times
q^4\sin^4\theta
\frac{\pi^4}{q^3|\Vec{k}-\Vec{q}|^3}
{\cal P}(q){\cal P}(|\Vec{k}-\Vec{q}|)
F(\Vec{k},\Vec{q},\eta')
F(\Vec{k},\Vec{q},\eta''),
\end{align}
where $\theta$ is the angle between $\Vec{q}$ and $\Vec{k}$.

%By an explicit calculation we obtain
%\begin{align}
%e^{(+)}_{ij}(\Vec{k})q^iq^j =
%\frac{1}{\sqrt{2}}q^2\sin^2\theta\cos(2\varphi),
%\quad
%e^{(-)}_{ij}(\Vec{k})q^iq^j =
%\frac{1}{\sqrt{2}}q^2\sin^2\theta\sin(2\varphi),
%\end{align}
%$\Vec{q}\cdot\Vec{k}=qk\cos\theta$,
%\begin{align}
%\int_0^{2\pi}\D\varphi\cos^2(2\varphi) =
%\int_0^{2\pi}\D\varphi\sin^2(2\varphi) = \pi = \frac{1}{2}\int_0^{2\pi}\D\varphi.
%\end{align}

The final step is to extract
from the above expression
the energy density parameter of GWs,
$\Omega_{\rm GW}(\eta, k)$, defined by
\begin{align}
\rho_{\rm GW}(\eta)=3\mpl^2H^2\int \Omega_{\rm GW}(\eta,k)\D\ln k.
\end{align}
In practice, $F$ depends on $\Vec{k}$, $\Vec{q}$, and $\eta$
through $q\eta$ and $|\Vec{k}-\Vec{q}|\eta$.
Therefore, it is convenient to introduce
dimensionless variables $u:=|\Vec{k}-\Vec{q}|/k$ and $v:=q/k$.
Using these variables,
the energy density parameter is expressed as
\begin{align}
\Omega_{\mathrm{GW}}(\eta,k)
&=\frac{1}{3H^2}\frac{1}{a^4(\eta)}
\int_0^\infty\D v \int^{1+v}_{|1-v|}\D u
\int^\eta_0\D\eta'\int^\eta_0\D\eta''
a(\eta')a(\eta''){\cal G}_k(\eta,\eta'){\cal G}_k(\eta, \eta'')
\notag \\ &\quad \times
\frac{k^4}{4}\frac{v^2}{u^2}\left[
1-\left(\frac{1+v^2-u^2}{2v}\right)^2
\right]^2
{\cal P}(ku){\cal P}(kv)
F(u,v,k\eta')
F(u,v,k\eta'')
\notag \\ &  =
\frac{k^2}{12{\cal H}^2}\int_0^\infty\D v\int_{|1-v|}^{1+v}\D u
 \frac{v^2}{u^2}\left[
1-\left(\frac{1+v^2-u^2}{2v}\right)^2
\right]^2
{\cal P}(ku){\cal P}(kv){\cal  I}^2(u,v,k\eta),
\end{align}
where
\begin{align}
{\cal I}(u,v,k\eta)=\frac{k}{a(\eta)}
\int_0^\eta \D\eta' a(\eta'){\cal G}_k(\eta,\eta')
F(u,v,k\eta').
\end{align}

The information of the initial conditions for scalar perturbations
is encoded in the power spectrum ${\cal P}$, while
the time evolution of the perturbations determines the form of the integral ${\cal I}$.
The two distinct effects are thus separated.
The gauge difference is essentially imprinted in ${\cal I}$.
This can be evaluated analytically for $w=0$ and $w=1/3$~\cite{Kohri:2018awv}
and numerically for the other values of $w$.

\section{Action Approach to Induced Gravitational Waves}

To derive the gauge-ready form of the source term for induced GWs
in a universe filled with an irrotational barotropic perfect fluid,
it is convenient to employ the action-based approach,
describing the fluid in terms of
a shift-symmetric k-essence field.

Our action is given by
\begin{align}
S=\int\D^4 x\sqrt{-g}\left[
\frac{\mpl^2}{2}R+P(X)
\right],\label{kessenceaction}
\end{align}
where $X:=-g^{\mu\nu}\partial_\mu\phi\partial_\nu\phi/2$.
The k-essence field is equivalent to a cosmological perfect fluid
whose energy density and pressure are given respectively by
$\rho=2XP_X-P$ and $p=P$. (Here and hereafter we write $\partial P/\partial X=P_X$.)
Therefore, a
$w=p/\rho=\;$const fluid can be mimicked by~\cite{Matarrese:1984zw,Garriga:1999vw}
\begin{align}
P\propto X^{(1+w)/2w}.
\end{align}
Hereafter we will assume that $w\ge 0$.
Note that the $w=0$ case appears to be singular,
but a careful inspection shows that the limit in fact
makes sense~\cite{Boubekeur:2008kn}.

The metric
in the $3+1$ Arnowitt-Deser-Misner (ADM) form
is given by
\begin{align}
\D s^2=-N^2\D t^2 +g_{ij}\left(\D x^i+N^i\D t\right)\left(\D x^j+N^j\D t\right).
\end{align}
We only fix the spatial coordinate, while leaving the temporal gauge
degree of freedom unfixed, as we are interested only in
the latter gauge difference. We therefore write the ADM variables
in terms of the scalar and tensor perturbations as
\begin{align}
N=1+\delta N,\quad N_i=\partial_i\chi,
\quad
g_{ij}=a^2e^{-2\psi}(e^h)_{ij},\label{pertmetric}
\end{align}
where $(e^h)_{ij}=\delta_{ij}+h_{ij}+h_{ik}h_{kj}/2+\cdots$.
The perturbed scalar field is written as
\begin{align}
 \phi = \bar\phi(t)+\delta\phi.\label{pertscalar}
\end{align}
We will omit the bar from the background value if unnecessary.
Using the temporal gauge degree of freedom one can
eliminate one of $\chi$, $\psi$, and $\delta \phi$.

The background equations are given by
\begin{align}
3\mpl^2H^2&=2XP_X-P,
\\
\mpl^2\left(3H^2+2\dot{H}\right)&=-P,
\\
\frac{\D}{\D t}\left(a^3\dot\phi P_X\right)
%(P_X+2XP_{XX})\ddot\phi+3H P_X\dot\phi
&=0.\label{phi_eom_0}
\end{align}

In order to derive the equations of motion for
the perturbations, we
substitute the metric~\eqref{pertmetric} and the
scalar field~\eqref{pertscalar} to the action~\eqref{kessenceaction}
and expand it to third order.
At third order
we only need the terms containing
one tensor and two scalars,
because the variation of such terms with respect to $h_{ij}$
leads to the source terms for GWs induced by scalar perturbations.
Thus, the action that suffices for our purpose is
\begin{align}
S=\int\D t \D^3x\left[
{\cal L}^{(2)}_s+{\cal L}_h^{(2)}+{\cal L}^{(3)}_{ssh}
\right],
\end{align}
where
\begin{align}
{\cal L}^{(2)}_s&=
a^3\mpl^2\left[
-3\dot\psi^2-\frac{\psi\partial^2\psi}{a^2}-2H\delta N\frac{\partial^2\chi}{a^2}
-2\dot\psi \frac{\partial^2\chi}{a^2}-6H\delta N\dot\psi+2\delta N\frac{\partial^2\psi}{a^2}
\right]
\notag \\ & \quad
+a^3 \left(XP_X+2X^2P_{XX}-3\mpl^2H^2\right)\delta N^2
+\frac{a^3}{2}(P_X+2XP_{XX})
\left(\dot{\delta\phi}^2-2\dot\phi \delta N\dot{\delta\phi}\right)
\notag \\ & \quad
-3a^3\dot\phi P_X\psi\dot{\delta\phi}
+\frac{aP_X}{2}\left(\delta\phi\partial^2\delta\phi
+2\dot\phi \delta\phi\partial^2\chi\right),
\label{Lag:linear order_s}
\\
{\cal L}^{(2)}_h&=
\frac{a^3\mpl^2}{8}\left[\dot h_{ij}^2-\frac{(\partial_k h_{ij})^2}{a^2}\right],
\label{Lag:linear order_h}
\end{align}
and
\begin{align}
{\cal L}^{(3)}_{ssh}&=
a\mpl^2\left[
2(H\delta N +\dot\psi)\chi_{,ij}h_{ij}
+\frac{1}{2}(\delta N+3\psi) \chi_{,ij}\dot h_{ij}
-\psi_{,i}\psi_{,j}h_{ij}+2\delta N_{,i}\psi_{,j}h_{ij}
+\frac{1}{4a^2}\partial^2\left(\chi_{,i}\chi_{,j}\right)h_{ij}
\right]
\notag \\ &\quad
+\frac{aP_X}{2}h_{ij}\left(\partial_i\delta\phi\partial_j\delta\phi
+2\dot\phi \partial_i\chi\partial_j\delta\phi\right).
\end{align}
The quadratic Lagrangian ${\cal L}_{s}^{(2)}$
yields the linearized equations of motion for
the scalar perturbations.
The variation of the above action with respect to $h_{ij}$
gives the equation of motion for $h_{ij}$
sourced by the scalar perturbations,
which takes the form of Eq.~\eqref{EoM of h genuniversel}.
Now it is straightforward to obtain
\begin{align}
S_{ij}(\eta,\Vec{x})&=\frac{1}{a^2}\partial^2(\chi_{,i}\chi_{,j})
+8\delta N_{(,i}\psi_{,j)}-4\psi_{,i}\psi_{,j}
+\frac{8}{a}({\cal H}\delta N+\psi')\chi_{,ij}
-\frac{2}{a^2}\frac{\D}{\D \eta}\left[a(\delta N+3\psi)\chi_{,ij}\right]
\notag \\ & \quad
+\frac{4XP_X}{\mpl^2H^2}\left[Q_{,i}Q_{,j}
+\frac{2{\cal H}}{a}\chi_{(,i}Q_{,j)}\right],
\label{def:S_ij}
\end{align}
where we defined $Q(t,\Vec{x}):=H\delta\phi/\dot\phi$.
Note that one can simplify the expression further by
using the background equation and write
$4XP_X/\mpl^2H^2=6(1+w)$.
This is the gauge-ready form of the source term for induced GWs.
Moving to the Fourier domain, we have
\begin{align}
S_A(\eta,\Vec{k})=e_{ij}^{(A)}(\Vec{k})\int \frac{\D^3q}{(2\pi)^{3/2}}q^iq^j
\left[
-\frac{k^2}{a^2}\chi(\eta,\Vec{q})\chi(\eta,\Vec{k}-\Vec{q})
+4\delta N(\eta,\Vec{q})\psi(\eta,\Vec{k}-\Vec{q})
+4\psi(\eta,\Vec{q})\delta N(\eta,\Vec{k}-\Vec{q})+\cdots
\right].\label{safourier}
\end{align}
This can be recast in the form of Eq.~\eqref{def:F}
by separating the scalar perturbations into
the transfer functions and the primordial amplitudes.

\section{The Gauge Dependence}

Now let us investigate the gauge dependence of induced GWs.
Specifically, we consider
the Newtonian gauge ($\chi=0$), the comoving gauge ($\delta\phi=0$),
and the uniform curvature gauge ($\zeta=0$).
The gauge dependence can be seen clearly by evaluating
the integral ${\cal I}(u,v,k\eta)$.
In radiation-dominated (RD) and matter-dominated (MD) universes
this can be done analytically.
For the other values of $w$, one needs to perform numerical integration
to evaluate ${\cal I}$ precisely.

\subsection{Newtonian Gauge}

We start with reproducing the standard Newtonian gauge result.
The Newtonian gauge is defined by
\begin{align}
\chi=0.
\end{align}
Following the conventional notation we write $\delta N= \Phi$ and $\psi=\Psi$.

From the Lagrangian~\eqref{Lag:linear order_s}
we obtain
\begin{align}
&\Psi''+2{\cal H}\Psi'+\frac{k^2}{3}(\Psi-\Phi)
+{\cal H}\Phi'+\frac{3}{2}(1-w){\cal H}^2\Phi
-\frac{3}{2}(1+w){\cal H}^2\left[\frac{Q'}{\cal H}+\frac{3}{2}(1-w)Q\right]=0,
\\
&-3{\cal H}\Psi'-k^2\Psi +\frac{3(1-w)}{2w}{\cal H}^2\Phi
-\frac{3(1+w)}{2w}{\cal H}^2\left[\frac{Q'}{\cal H}+\frac{3}{2}(1-w)Q\right]=0,
\\
&Q''+{\cal H}\left(2Q'-\Phi'-3w\Psi'\right)+w k^2Q =0,
\end{align}
where
we moved to the Fourier domain and used the conformal time.
The solution to the above set of equations is given by
\begin{align}
\Phi=\Psi&=A_\Phi(\Vec{k})f_\Phi(\eta, k),\label{newton_phi_psi}
\\
Q&= \frac{A_\Phi(\Vec{k})}{3(1+w)}\left[2f_\Phi+(1+3w)\eta f_\Phi'\right],
\label{newton_Q}
\end{align}
where
\begin{align}
f_\Phi(\eta,k):=\Gamma(\nu+2)\left(\frac{\sqrt{w}k\eta}{2}\right)^{-\nu-1}
J_{\nu+1}(\sqrt{w}k\eta),
\end{align}
$A_\Phi(\Vec{k})$ is the amplitude of $\Phi$ at $\eta=0$,
and we discarded another independent solution that diverges at $\eta=0$.
We thus have $\Phi =\Psi = A_\Phi$ and $Q= 2A_\Phi/[3(1+w)]$
at $\eta=0$.
Substituting the above result to the source term~\eqref{safourier},
one can compute $F(u,v,k\eta)$.

In a RD universe,
we find
\begin{align}
\frac{1}{54}F_{{\rm RD},\chi}(u,v,k\eta)
&=\frac{18}{u^2v^2k^4\eta^4}\cos\left(\frac{uk\eta}{\sqrt{3}}\right)
\cos\left(\frac{vk\eta}{\sqrt{3}}\right)
\notag \\ &\quad
+
\frac{2\sqrt{3}}{u^3v^2k^5\eta^5}
\left(
u^2k^2\eta^2-9\right)
\sin\left(\frac{uk\eta}{\sqrt{3}}\right)
\cos\left(\frac{vk\eta}{\sqrt{3}}\right)
\notag \\ &\quad
+
\frac{2\sqrt{3}}{u^2v^3k^5\eta^5}
\left(
v^2k^2\eta^2-9\right)
\cos\left(\frac{uk\eta}{\sqrt{3}}\right)
\sin\left(\frac{vk\eta}{\sqrt{3}}\right)
\notag \\ & \quad
+\frac{1}{u^3v^3k^6\eta^6}
\left[
54-6(u^2+v^2)k^2\eta^2+u^2v^2k^4\eta^4
\right]\sin\left(\frac{uk\eta}{\sqrt{3}}\right)
\sin\left(\frac{vk\eta}{\sqrt{3}}\right).\label{fradnewton}
\end{align}
Following Ref.~\cite{Kohri:2018awv} one can evaluate analytically the
integral ${\cal I}(u,v,k\eta)$ for $k\eta \gg 1$:
\begin{align}
{\cal I}_{{\rm RD},\chi}(u,v,k\eta)
=\frac{1}{4k\eta}\left[{\cal I}_1\cos(k\eta)+{\cal I}_2\sin(k\eta)\right],
\end{align}
where
\begin{align}
{\cal I}_1&:= \frac{27}{2}\frac{u^2+v^2-3}{u^3v^3}\left[-4uv+
(u^2+v^2-3)\ln\left|\frac{3-(u+v)^2}{3-(u-v)^2}\right|
\right],
\\
{\cal I}_2&:=\frac{27}{2}\pi
\frac{(u^2+v^2-3)^2}{u^3v^3}\Theta(u+v-\sqrt{3}),
\end{align}
with $\Theta$ being the step function.
Its oscillation average is therefore
\begin{align}
\overline{{\cal I}^2_{{\rm RD},\chi}}=\frac{{\cal I}_1^2+{\cal I}_2^2}{32(k\eta)^2}.
\end{align}

In a MD universe,
we take the $w\to 0$ limit in Eqs.~\eqref{newton_phi_psi} and~\eqref{newton_Q}
and obtain
$\Phi=\Psi =A_\Phi$,
$Q=(2/3)A_\Phi$.
Thus, it is easy to see
\begin{align}
F_{{\rm MD},\chi}=\frac{20}{3},
\end{align}
and thus
\begin{align}
{\cal I}_{{\rm MD},\chi}=
\frac{20[k\eta-\sin(k\eta)]}{(k\eta)^2}.
\end{align}

\subsection{Comoving Gauge v.s. Newtonian Gauge}

The comoving gauge is defined by
\begin{align}
\delta\phi=0.
\end{align}
We write the comoving curvature perturbation as $\zeta = -\psi$.

The linear equations of motion in the comoving gauge are
\begin{align}
&6{\cal H}\zeta' +2k^2\zeta+\frac{2}{a}{\cal H}k^2\chi
+\frac{3(1-w)}{w}{\cal H}^2\delta N=0,
\\
&\delta N = \frac{\zeta'}{{\cal H}},
\\
&
\zeta''+2{\cal H}\zeta'+\frac{k^2}{3}\zeta +\frac{k^2}{3a}\left(\chi'
+{\cal H} \chi\right)-{\cal H}\delta N'+\frac{k^2}{3} \delta N
-\frac{3}{2}(1-w){\cal H}^2\delta N =0.
\end{align}
The solution regular at $\eta=0$ is given by
\begin{align}
\zeta = A_\zeta(\Vec{k})f_\zeta(k,\eta),
\quad
\delta N= \frac{1+3w}{2}A_\zeta(\Vec{k})\eta f_\zeta',
\quad
\frac{k}{a}\chi &=-\frac{A_\zeta(\Vec{k})}{2}
\left[\frac{3(1+w)}{w}\frac{f_\zeta'}{k}+(1+3w)k\eta f_\zeta\right],
\label{ssol_comov}
\end{align}
where
\begin{align}
f_\zeta(k,\eta):=\Gamma(\nu+1)\left(\frac{\sqrt{w}k\eta}{2}\right)^{-\nu}
J_\nu(\sqrt{w}k\eta).\label{def:f_zeta}
\end{align}
It is well-known that the primordial amplitude in the comoving gauge
is related to that in the Newtonian gauge by
\begin{align}
A_\zeta(\Vec{k})=-\frac{5+3w}{3(1+w)}A_\Phi(\Vec{k}).
\label{phi-zeta-rel}
\end{align}

To understand the gauge dependence in a RD universe,
it will be helpful to see the behavior of
the scalar perturbations for $k\eta\gg 1$. For $w=1/3$, we have
\begin{align}
\zeta =A_\zeta \frac{\sin(k\eta/\sqrt{3})}{k\eta/\sqrt{3}},
\quad
\delta N \approx A_\zeta\cos(k\eta/\sqrt{3}),
\quad
\frac{k}{a}\chi\approx -\sqrt{3}A_\zeta \sin(k\eta/\sqrt{3}).
\end{align}
This should be contrasted with the behavior of
the Newtonian gauge perturbations in a RD universe for $k\eta\gg 1$:
\begin{align}
\Phi=\Psi \approx -9A_\Phi \frac{\cos(k\eta/\sqrt{3})}{k^2\eta^2},
\quad
Q\approx \frac{3}{2}A_\Phi \frac{\sin(k\eta/\sqrt{3})}{k\eta/\sqrt{3}}.
\end{align}
We see that in the comoving gauge the source term contains
the terms that do not decay at late times.
Accordingly, we have
\begin{align}
F_{{\rm RD},\delta\phi}(u,v,k\eta)
&=\left\{-2-\frac{12}{u^2v^2k^2\eta^2}
\left[
3-2\left(u^2+v^2\right)
\right]
\right\}\cos\left(\frac{uk\eta}{\sqrt{3}}\right)
\cos\left(\frac{vk\eta}{\sqrt{3}}\right)
\notag \\ &\quad
+
\frac{2\sqrt{3}}{u^3v^2k^3\eta^3}
\left\{
6\left[3-2(u^2+v^2)\right]
+u^2\left(-3+u^2+2v^2\right)k^2\eta^2
\right\}
\sin\left(\frac{uk\eta}{\sqrt{3}}\right)
\cos\left(\frac{vk\eta}{\sqrt{3}}\right)
\notag \\ &\quad
+
\frac{2\sqrt{3}}{u^2v^3k^3\eta^3}
\left\{
6\left[3-2(u^2+v^2)\right]
+v^2\left(-3+2u^2+v^2\right)k^2\eta^2
\right\}
\cos\left(\frac{uk\eta}{\sqrt{3}}\right)
\sin\left(\frac{vk\eta}{\sqrt{3}}\right)
\notag \\ & \quad
+\frac{1}{u^3v^3k^4\eta^4}
\biggl\{
-36\left[3-2(u^2+v^2)\right]
\notag \\ & \qquad
+k^2\eta^2\left[6(u^2+v^2)-k^2\eta^2u^2v^2\right]
\left[3-(u^2+v^2)
\right]
\biggr\}\sin\left(\frac{uk\eta}{\sqrt{3}}\right)
\sin\left(\frac{vk\eta}{\sqrt{3}}\right),
\end{align}
which does not decay at late times, in contrast to
the Newtonian gauge result~\eqref{fradnewton}.
It then follows that
\begin{align}
{\cal I}_{{\rm RD},\delta\phi}
&=\left(\frac{2}{3}\right)^{2}I_{{\rm RD},\chi}
-\frac{3}{2u^2v^2k\eta }\left\{
3(2u^2-3uv+2v^2)\cos\left[(u-v)\frac{k\eta}{\sqrt{3}}\right]
+3(2u^2+3uv+2v^2)\cos\left[(u+v)\frac{k\eta}{\sqrt{3}}\right]
\right\}
\notag
\\ & \quad
+\frac{\sqrt{3}}{2uv}\left\{
(u-v)\sin\left[(u-v)\frac{k\eta}{\sqrt{3}}\right]
-(u+v)\sin\left[(u+v)\frac{k\eta}{\sqrt{3}}\right]
\right\},\label{result_rd_com}
\end{align}
where we took the limit $k\eta\gg1$.
If we had only the first term in Eq.~\eqref{result_rd_com},
the energy density of induced GWs would always be gauge-invariant,
as the factor $(2/3)^2$ is canceled
in the final result due to the relation~\eqref{phi-zeta-rel}.
The first line decays as $\sim \eta^{-1}$, while the
second line just oscillates, and hence the latter in fact dominates
at late times, resulting in a large gauge dependence.
This is essentially due to the first term in the source~\eqref{def:S_ij}.

{The large gauge dependence we have observed is in fact generic to
the other values of $w \,(>0)$. In the Newtonian gauge, we have, for $k\eta\gg 1$,
\begin{align}
\Phi=\Psi \sim \eta^{-\nu-3/2},\quad Q\sim \eta^{-\nu-1/2}.
\end{align}
However, in the comoving gauge we have
\begin{align}
\zeta\sim  \eta^{-\nu-1/2},\quad \delta N\sim \frac{k}{a}\chi
\sim \eta^{-\nu+1/2},
\end{align}
which shows that the first term in the source~\eqref{def:S_ij}
always overwhelms the other contributions and causes
a large gauge dependence.

Let us then consider a MD universe.
Since one has $w$ in the denominator in Eq.~\eqref{ssol_comov},
the $w\to 0$ limit in the k-essence description of a fluid
seems particularly subtle in the comoving gauge. However,
for $w\ll 1$, $f_\zeta$ is approximated by
\begin{align}
f_\zeta = 1-\frac{w}{10}k^2\eta^2,
\end{align}
and using this one finds
\begin{align}
\zeta = A_\zeta,\quad \delta N=0,\quad
\frac{k}{a}\chi = -\frac{1}{5}A_\zeta k\eta
\end{align}
in the $w\to 0$ limit~\cite{Boubekeur:2008kn}.
Thus, one can safely take the $w\to0$ limit.
Note that $(k/a)\chi$ grows in time.
This is again different from the behavior of the Newtonian gauge variables
in a MD universe: $\Phi$, $\Psi$, and $Q$ remain constant.
This difference gives rise to a growing contribution in $F$ and ${\cal I}$:
\begin{align}
F_{{\rm MD},\delta\phi}
&=2-\frac{k^2\eta^2}{25},
\\
{\cal I}_{{\rm MD},\delta\phi}
&=\left(\frac{3}{5}\right)^2{\cal I}_{{\rm MD},\chi}-\frac{k\eta}{5}.
\label{Imd_comv}
\end{align}
If we had only the first term in Eq.~\eqref{Imd_comv},
there would be no gauge dependence in induced GWs, given that
Eq.~\eqref{phi-zeta-rel} accounts for the factor $(3/5)^2$.
However, this term decays as $\sim \eta^{-1}$,
and so the second term dominates at late times.
Therefore, there is a large gauge dependence also in this case.
Again, this is caused by the first term in the source~\eqref{def:S_ij}.

\subsection{Uniform Curvature Gauge v.s. Newtonian Gauge}

Finally, let us consider the
uniform curvature gauge defined by
\begin{align}
\psi=0.
\end{align}

The evolution of the scalar perturbations
in the uniform curvature gauge is governed by
\begin{align}
&
{\cal H}Q'+\frac{3}{2}(1-w){\cal H}^2Q - \frac{2w}{3(1+w)}\frac{{\cal H}}{a} k^2\chi
-\frac{1-w}{1+w}{\cal H}^2\delta N=0,
\\
&
\delta N =\frac{3(1+w)}{2}Q,
\\
& Q''+{\cal H}(2Q'-\delta N')+wk^2 \left(Q+\frac{{\cal H}}{a}\chi\right)=0.
\end{align}
The non-decaying solution is given by
\begin{align}
 Q =A_Q(\Vec{k})f_Q(\eta,k),
 \quad
 \delta N = \frac{3(1+w)}{2}A_Q(\Vec{k}) f_Q,
 \quad
 \frac{k}{a}\chi = \frac{3(1+w)}{2w}A_Q(\Vec{k})\frac{f_Q'}{k},
\end{align}
where
\begin{align}
  f_Q(\eta,k)=\Gamma(\nu+1)\left(\frac{\sqrt{w}k\eta}{2}\right)^{-\nu}
  J_\nu(\sqrt{w}k\eta).
\end{align}
This function is the same as Eq.~\eqref{def:f_zeta}.
The primordial amplitude $A_Q$ is related to $A_\zeta$ (and $A_\Phi$)
by
\begin{align}
A_Q(\Vec{k})=-A_\zeta(\Vec{k})=\frac{5+3w}{3(1+w)}A_\Phi(\Vec{k}).
\label{Q-zeta-rel}
\end{align}

In a RD universe, we have
\begin{align}
Q=\frac{\delta N}{2}=A_Q \frac{\sin(k\eta/\sqrt{3})}{k\eta/\sqrt{3}},
\quad
\frac{k}{a}\chi \approx 6A_Q\frac{\cos(k\eta/\sqrt{3})}{k\eta}
\end{align}
for $k\eta \gg 1$. Therefore,
unlike in the comoving gauge,
the perturbations in the uniform curvature gauge decay as $\sim \eta^{-1}$.
It is then straightforward to compute
\begin{align}
F_{{\rm RD},\zeta}(\Vec{k},\Vec{q},\eta)
&=\frac{12(u^2+v^2-3)}{u^2v^2k^2\eta^2}
\cos\left(\frac{u k\eta}{\sqrt{3}}\right)\cos\left(\frac{v k\eta}{\sqrt{3}}\right)
-\frac{12\sqrt{3}(u^2+v^2-3)}{u^3v^2k^3\eta^3}
\sin\left(\frac{u k\eta}{\sqrt{3}}\right)\cos\left(\frac{v k\eta}{\sqrt{3}}\right)
\notag \\ &\quad
-\frac{12\sqrt{3}(u^2+v^2-3)}{u^2v^3k^3\eta^3}
\cos\left(\frac{u k\eta}{\sqrt{3}}\right)\sin\left(\frac{v k\eta}{\sqrt{3}}\right)
+\frac{36 (u^2+v^2-3)}{u^3v^3k^4\eta^4}
\sin\left(\frac{u k\eta}{\sqrt{3}}\right)\sin\left(\frac{v k\eta}{\sqrt{3}}\right).
\end{align}
This expression is clearly different from
the Newtonian gauge result~\eqref{fradnewton}.
However, integrating this to get ${\cal I}$ we find
\begin{align}
{\cal I}_{{\rm RD},\psi} = \left(\frac{2}{3}\right)^2 {\cal I}_{{\rm RD},\chi}.
\label{Ird_uc}
\end{align}
Taking into account the relation~\eqref{Q-zeta-rel},
we see from Eq.~\eqref{Ird_uc}
that the comoving gauge and the uniform curvature gauge
give the identical result on the energy density of induced GWs.

To see whether this is accidental or not,
we evaluate ${\cal I}$ numerically for
the other values of $w\,(\lesssim 1)$.
Examples of our numerical investigation
are presented in Figs.~\ref{fig:1}--\ref{fig:3}.
In Figs.~\ref{fig:1} and~\ref{fig:2}
we present the comparison of ${\cal I}$
in the Newtonian and uniform curvature gauges
for different values of $w$ with $u$, $v$, $k\eta$ being fixed.
We also show in Fig.~\ref{fig:3}
the comparison of ${\cal I}$
as a function of $k\eta$ for $w=2/3$.
These results imply that the following relation holds:
\begin{align}
{\cal I}_{,\psi}=\left[\frac{3(1+w)}{5+3w}\right]^2{\cal I}_{,\chi}.
\end{align}
We thus conclude that the Newtonian gauge and the uniform curvature
gauge give the identical result on $\Omega_{\rm GW}$ for $w>0$.

%%%%%%%%%%%%%%%%%%%%%%%%%
\begin{figure}[tb]
    \includegraphics[keepaspectratio=true,height=80mm]{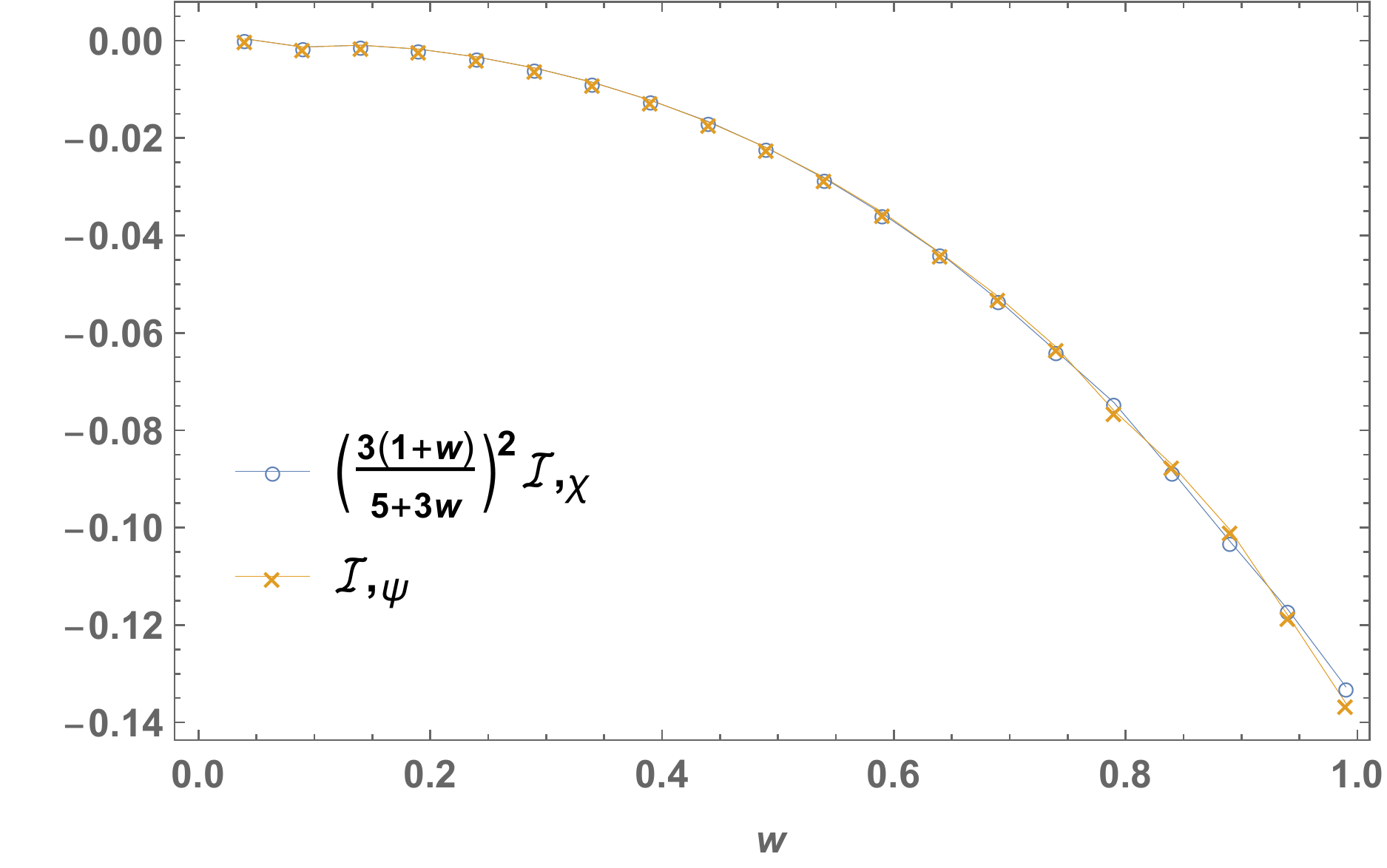}
     \caption{Comparison of ${\cal I}(2,2,250)$ (as a function of $w$) computed
     in the Newtonian and uniform curvature gauges.
	}
     \label{fig:1}
\end{figure}
%%%%%%%%%%%%%%%%%%%%%%%%%
%%%%%%%%%%%%%%%%%%%%%%%%%
\begin{figure}[tb]
    \includegraphics[keepaspectratio=true,height=80mm]{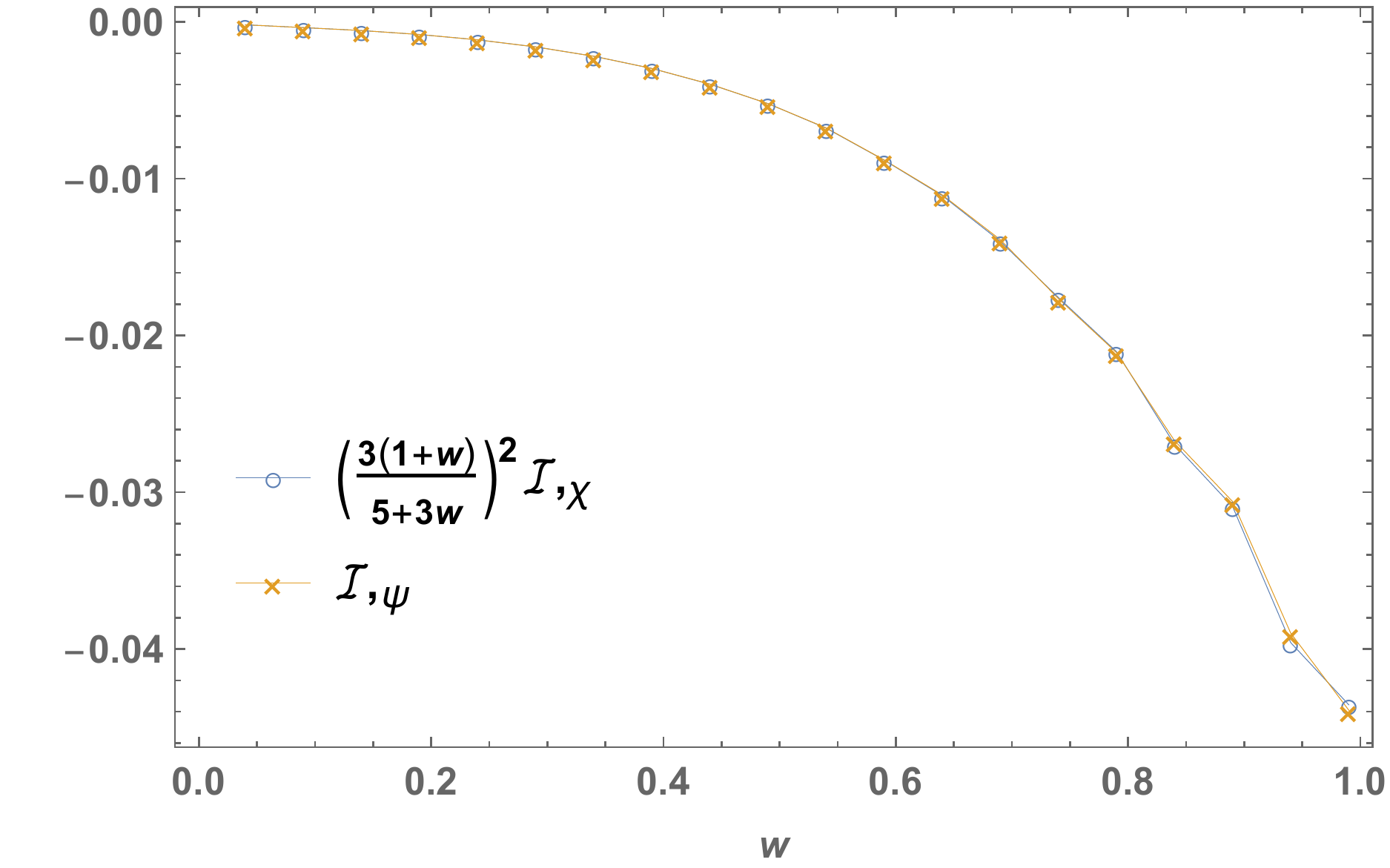}
     \caption{Comparison of ${\cal I}(5,5.5,250)$ (as a function of $w$) computed
     in the Newtonian and uniform curvature gauges.
	}
     \label{fig:2}
\end{figure}
%%%%%%%%%%%%%%%%%%%%%%%%%
%%%%%%%%%%%%%%%%%%%%%%%%%
\begin{figure}[tb]
    \includegraphics[keepaspectratio=true,height=80mm]{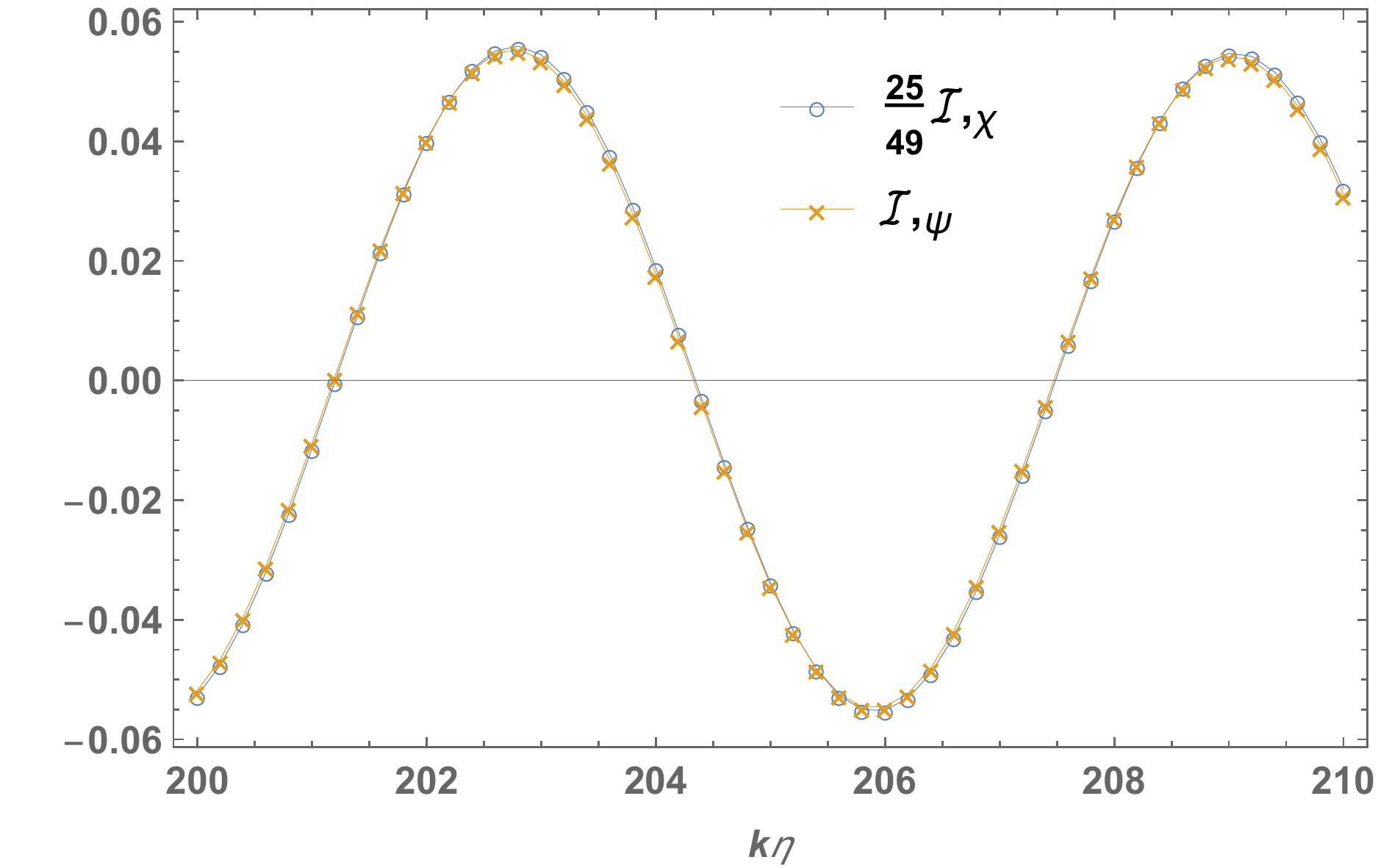}
     \caption{Comparison of ${\cal I}(2,2,k\eta)$ computed
     in the Newtonian and uniform curvature gauges for $w=2/3$.
	}
     \label{fig:3}
\end{figure}
%%%%%%%%%%%%%%%%%%%%%%%%%

This is, however, not true in the case of $w=0$.
Again, there is a subtlety regarding $w$
in the denominator, but this can be circumvented
in the same way as in the comoving gauge.
Since $f_Q= 1 -wk^2\eta^2/10$ for $w\ll 1$,
we have
\begin{align}
\frac{k}{a}\chi = -\frac{3}{10}A_Q k\eta.
\end{align}
Therefore, the evolution of the scalar perturbations
in a MD universe
is very similar to that in the comoving gauge.
This yields the following result on $F$ and ${\cal I}$:
\begin{align}
F_{{\rm MD},\psi}=\frac{3}{2}-\frac{9}{100}k^2\eta^2,
\quad
{\cal I}_{{\rm MD},\psi}
&=\left(\frac{3}{5}\right)^2{\cal I}_{{\rm MD},\chi}-\frac{9}{20}k\eta.
\end{align}
Therefore, in a MD universe,
the uniform curvature gauge and the Newtonian gauge
give very different results. Rather, the uniform curvature gauge
is similar to the comoving gauge
and their results are different only by a factor at late times:
${\cal I}_{{\rm MD},\psi}\approx (9/4){\cal I}_{{\rm MD},\delta\phi}$.
Here again, this is due to the first term in the source~\eqref{def:S_ij}.
The factor $9/4$ comes from the coefficients of the solution of $\chi$:
$(3/10)^2=(9/4)\times (1/5)^2$.
It is exactly this factor that explains
the difference between the comoving and uniform curvature gauges
in the numerical calculation in~\cite{Hwang:2017oxa}.

\section{Concluding Remarks}

In this paper, we have revisited the issue of the gauge dependence of gravitational waves
(GWs) induced at second order from scalar perturbations.
We have evaluated the energy density of induced GWs in different gauges
in a universe dominated by a perfect fluid whose equation-of-state
parameter $w$ is constant, and arrived at the following conclusions:
(i) the amplitude of induced GWs in the comoving gauge is
significantly larger than that in
the Newtonian gauge for any $w\,(\ge 0)$,
and this huge gauge dependence is a consequence of the presence of
the shift vector; (ii) for $w>0$ the Newtonian gauge result agrees
with that of the uniform curvature gauge; (iii)
for $w=0$ the uniform curvature gauge result differs only by a factor
from that of the comoving gauge, but deviates significantly from that of
the Newtonian gauge.
Our calculation has been done analytically for $w=0$ and $w=1/3$
using the method of Ref.~\cite{Kohri:2018awv}.
The above conclusions are
consistent with the previous numerical result~\cite{Hwang:2017oxa}.
The gauge dependence has been clarified based only on the evolution of
the perturbations, and hence our result is robust against
the input form of the primordial power spectrum of the scalar perturbations.

For simplicity and clarity, we have focused on the ideal case with
$w=\;$const rather than the realistic and conventional cosmological setup.
Nevertheless, we believe that
the present paper would be of help to gaining a deeper understanding
of the gauge dependence of scalar-induced GWs.

As was noted in Ref.~\cite{Hwang:2017oxa}, the appropriate gauge
one should choose depends on what quantity one measures in each observation.
Given that there is a large gauge dependence of second-order GWs,
it would be important to address this issue and identify the true observables.

\vspace{5mm}
{\bf Note added:}
While we were in the final stage of this work, the paper
by J. O. Gong~\cite{Gong:2019mui}
appeared in the arXiv, where the gauge dependence of induced gravitational
waves was studied analytically by comparing the Newtonian and comoving gauge
results in a matter-dominated universe.
Our conclusion agrees with his where we overlap.

%--- Acknowledgements ---%--- Acknowledgements ---%--- Acknowledgements ---%
\acknowledgments
We are grateful to K. Inomata and R. Saito
for fruitful discussions.
The work of KT was supported by the Rikkyo University Special Fund for Research.
The work of TK was supported by
MEXT KAKENHI Grant Nos.~JP15H05888, JP17H06359, JP16K17707, and JP18H04355.
%--- Acknowledgements ---%--- Acknowledgements ---%--- Acknowledgements ---%

%\appendix

%-----------------------------------------------------------------%
\bibliographystyle{JHEP.bst}
\bibliography{refs}
%-----------------------------------------------------------------%
\end{document}